\documentclass[journal,12pt,onecolumn, draftclsnofoot]{IEEEtran}
\usepackage[dvips]{graphicx}
\usepackage{multirow}
\usepackage{caption}
\usepackage{multicol,lipsum}
\usepackage{mathtools, cuted}
\usepackage{comment}
\usepackage{amsmath}
\usepackage{amsthm}
\usepackage[flushleft]{threeparttable}
\usepackage{array}
\usepackage{booktabs}
\usepackage{dirtytalk}
\newcommand{\PreserveBackslash}[1]{\let\temp=\\#1\let\\=\temp}
\newcolumntype{C}[1]{>{\PreserveBackslash\centering}p{#1}}
\newcolumntype{R}[1]{>{\PreserveBackslash\raggedleft}p{#1}}
\newcolumntype{L}[1]{>{\PreserveBackslash\raggedright}p{#1}}

\usepackage{bm}
\usepackage{graphicx,subfigure}
\usepackage{algorithmic, algorithm}
\usepackage{cite}
\usepackage{amsmath}
\usepackage{amssymb}
\usepackage{psfrag}
\usepackage{empheq}
\usepackage{dirtytalk}
\usepackage{latexsym, amsmath, subfigure, color, amsfonts, amssymb,graphicx}
\usepackage{wrapfig} 

\IEEEoverridecommandlockouts
\usepackage{tabularx}
\makeatletter
\def\hlinewd#1{%
\noalign{\ifnum0=`}\fi\hrule \@height #1 %
\futurelet\reserved@a\@xhline}
\makeatother
\begin{document}
\title{Pruning the Pilots: Deep Learning-Based Pilot Design and Channel Estimation for MIMO-OFDM Systems}
\author{Mahdi Boloursaz Mashhadi and Deniz G\"{u}nd\"{u}z\\
\IEEEauthorblockA{Dept. of Electrical and Electronic Eng., Imperial College London, UK\\
Email: \{m.boloursaz-mashhadi, d.gunduz\}@imperial.ac.uk}}

\maketitle
\begin{abstract}
With the large number of antennas and subcarriers the overhead due to pilot transmission for channel estimation can be prohibitive in wideband massive multiple-input multiple-output (MIMO) systems. This can degrade the overall spectral efficiency significantly, and as a result, curtail the potential benefits of massive MIMO. In this paper, we propose a neural network (NN)-based joint pilot design and downlink channel estimation scheme for frequency division duplex (FDD) MIMO orthogonal frequency division multiplex (OFDM) systems. The proposed NN architecture uses fully connected layers for frequency-aware pilot design, and outperforms linear minimum mean square error (LMMSE) estimation by exploiting inherent correlations in MIMO channel matrices utilizing convolutional NN layers. \color{black}Our proposed NN architecture uses a non-local attention module to learn longer range correlations in the channel matrix to further improve the channel estimation performance.\color{black} We also propose an effective pilot reduction technique by gradually pruning less significant neurons from the dense NN layers during training. This constitutes a novel application of NN pruning to reduce the pilot transmission overhead. Our pruning-based pilot reduction technique reduces the overhead by allocating pilots across subcarriers non-uniformly and exploiting the inter-frequency and inter-antenna correlations in the channel matrix efficiently through convolutional layers and attention module.

\makeatletter{\renewcommand*{\@makefnmark}{}\footnotetext{This work was supported by the European Research Council (ERC) through project BEACON (grant no 677854).}\makeatother}
\end{abstract}
\section{Introduction}\label{intro}
Massive multiple-input multiple-output (MIMO) systems are considered as the main enabler of 5G and future wireless networks thanks to their ability to serve a large number of users simultaneously, achieving impressive levels of spectral efficiency. A base station (BS) with a massive number of antennas relies on accurate downlink channel state information (CSI) to achieve the promised performance gains. Therefore, massive MIMO systems are more amenable to time division duplex (TDD) operation, which, thanks to the reciprocity of the uplink and downlink channels, does not require downlink channel estimation at the users. FDD operation is more desirable due its improved coverage and reduced interference; however, channel reciprocity does not hold in FDD. In FDD MIMO, the BS broadcasts downlink pilot signals, the users estimate the channel from the received pilots and transmit the CSI feedback to the BS. The resulting overhead becomes significant due to the large number of antennas and users; and hence, efficient pilot design and channel estimation are crucial to reduce the overhead.

In massive MIMO systems where the pilot length is typically much smaller than the number of antennas, channel estimation becomes severely underdetermined. Hence, simple least squares (LS) or linear minimum mean square error (LMMSE) channel estimation and orthogonal FFT pilots perform poorly. To estimate the channel more efficiently and reduce the pilot overhead, many previous works take a model-based estimation approach assuming sparse \cite{CSest1, CSest2, CSest3, CSest4, CSest5} or low-rank \cite{LRest1, LRest2} models on the channel matrix and utilize compressive sensing (CS)-based reconstruction techniques to estimate the channel or design improved pilot sequences. CS-based approaches rely on sparse or low-rank properties of the channel, and do not take into account the inherent statistical correlations and structures beyond sparse or low-rank patterns. Moreover, CS-based reconstruction techniques employ computationally demanding iterative algorithms, which imposes an additional burden on the users.

More recently, deep learning (DL)-based approaches have been used for massive MIMO CSI acquisition and showed significant improvements in comparison with their counterparts based on sparsity and compressive sensing (refer to \cite{overview1, overview2}, and references therein). In these works, neural network (NN) architectures are trained over large CSI datasets to learn complex distributions, structures, and correlations, and exploit them for data-driven pilot design \cite{EstPil}, channel estimation \cite{DLest1, DL1Bit, balevi2019twostage, DLFewBit, 2018}, compression \cite{CSINET, DLCSI2, DLCSI4, CSINETPlus, yang2019deepJ, Cooperative} and feedback \cite{FEDDEL, mashhadi2019cnnbased}. Many of these works focus on a single task and propose a NN architecture to achieve optimized performance for it. While designing a single NN architecture to simultaneously handle all or several of these tasks is desirable for an end-to-end optimized performance, the resulting NN may be more complex, and require a longer training process. In this paper, we consider joint pilot design and channel estimation for downlink FDD massive MIMO systems.  

In \cite{Navid, DLest1}, the authors proposed a convolutional neural network (CNN)-based structure for massive MIMO channel estimation. Their proposed architecture outperforms non-ideal LMMSE-based channel estimation (where the required covariance matrices are estimated from a coarse initial estimate of the channel at the receiver) and approach ideal LMMSE (with perfect knowledge of the covariance matrices assumed at the receiver).  In \cite{EstPil}, the authors use dense layers (which represent the pilots) followed by subsequent convolutional layers for joint pilot design and channel estimation. However, they design the same pilots for all subcarriers, which not only neglects frequency specific features in the CSI, but also results in a large PAPR, which is practically undesirable.

In this paper, we propose a NN-based scheme consisting of convolutional and dense layers for downlink pilot design and channel estimation in FDD massive MIMO-OFDM systems. Our proposed NN-based scheme exploits frequency-specific features in a data-driven manner for more efficient pilot design and channel estimation. The specific contributions of this paper can be summarized as follows: 
\begin{itemize}
    \item We propose a NN architecture which learns MIMO channel statistics over a dataset during training and exploits it to jointly design pilots and estimate the channel without requiring covariance matrices or other prior statistical assumptions on the channel matrix. By joint optimization of the pilot signals and the channel estimator, our proposed NN structure improves the normalized mean square error (NMSE) in comparison with \cite{Navid, DLest1}, which uses simple FFT pilots with NN-based channel estimation.
    
    \color{black}\item Our proposed NN architecture uses dense layers to design frequency-aware pilot signals followed by convolutional layers to learn the inherent correlations in the MIMO-OFDM channel, and to exploit them for efficient channel estimation. We also use a non-local attention module, which enables the NN to learn and exploit longer range correlations in the channel matrix.\color{black}

    \item Our proposed frequency-aware NN-based pilot design scheme outperforms the simple frequency-independent pilot scheme used in \cite{EstPil} by a large margin. This is due to the fact that our NN-based structure learns frequency-specific features of the channel matrices over the training dataset and exploits them to optimize pilots over different frequencies in a data-driven manner. Frequency-aware pilot design also enables us to allocate pilots non-uniformly over different subcarriers by our proposed pilot reduction technique based on NN pruning.

    \item We propose a pilot reduction technique based on NN pruning, which effectively reduces the pilot overhead by allocating pilots across subcarriers non-uniformly; fewer pilots are transmitted on subcarriers that can be satisfactorily reconstructed by the subsequent convolutional layers utilizing inter-frequency correlations. According to the simulation results, this scheme effectively improves the estimation NMSE for the same amount of time-frequency resources allocated uniformly over subcarriers for pilot transmission.

    \item We show by extensive simulations that our proposed scheme significantly outperforms the ideal LMMSE channel estimation as well as the recent works in \cite{Navid, DLest1, EstPil}. We provide an ablation study to investigate the performance of NN-based pilot design and channel estimation separately.  
\end{itemize}

The rest of this paper is organized as follows. In Section II, we present the system model. In Section III we present the proposed NN architecture for joint pilot design and channel estimation. In Section IV we present our pilot reduction scheme by NN pruning. Section V provides the simulation results, and Section VI concludes the paper.

\section{System Model}
\label{sec1}

We consider an FDD massive MIMO system, where a BS with $N$ antennas serves a single-antenna user utilizing orthogonal frequency division multiplexing (OFDM) over $M$ subcarriers. We denote the downlink channel by $\mathbf{H}= [\mathbf{h}_1, \mathbf{h}_2, \hdots, \mathbf{h}_{M}] \in \mathbb{C}  ^{N \times M}$, where $\mathbf{h}_m \in \mathbb{C}^N$ is the channel gain vector over subcarrier $m$, for $m= 1, \ldots, M$. We assume that the BS is equipped with a uniform linear array (ULA) with response vector:
\small
\[
\mathbf{a}(\phi)=\frac{1}{\sqrt{N}}[1, e^{-j\frac{2\pi d}{\lambda} \sin{\phi}}, \cdots, e^{-j\frac{2\pi d}{\lambda} (N-1) \sin{\phi}}]^T,
\]
\normalsize
where  $\phi$ is the angle of departure (AoD), and $d$ and $\lambda$ denote the distance between adjacent antennas and carrier wavelength, respectively. The channel gain is a summation of multipath components \cite{channel} given by 
\small
\begin{align}\label{Hform}
    \mathbf{h}_m=\sqrt{\frac{N}{P}}\sum^{P}_{p=1} \alpha_p e^{-j2\pi \tau_p f_s \frac{m}{M}} \mathbf{a}(\phi),
\end{align}
\normalsize
where $P$ is the number of multipath components, $f_s$ is the sampling rate, $\tau_p$ is the delay, and $\alpha_p$ is the propagation gain of the $p^{th}$ path. According to Eq. (\ref{Hform}), entries of the channel matrix $\mathbf{H}$ are correlated for nearby sub-carriers and antennas due to similar propagation paths, gains, and AoDs/AoAs. There also exist inherent characteristics in MIMO environments due to specific user distributions, scattering parameters, geometry, etc., that cause common structures among MIMO channel matrices.

\begin{figure}[t!]
\centering
\includegraphics[scale=0.25]{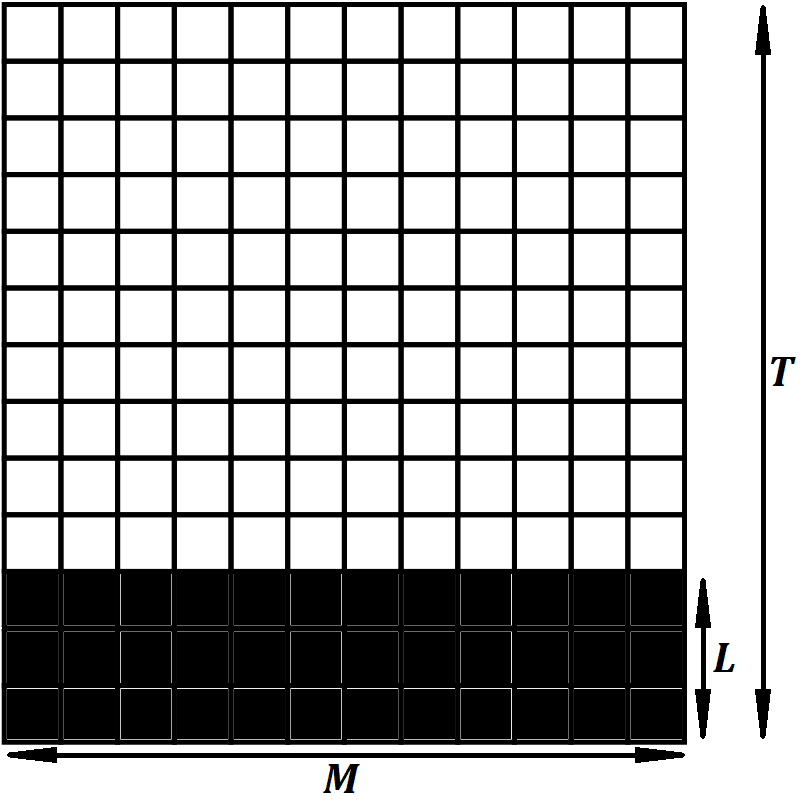}
\caption{\color{black}The time-frequency resource grid structure.}
\label{RG}
\end{figure}

\color{black}Fig. \ref{RG} depicts the time-frequency resource grid structure, where we use a pilot block of size $L\times M$ (denoted by the black slots in Fig. \ref{RG}) to estimate the channel over a time-frequency grid of size $T \times M$ and $L \leq T$. The coherence time of the channel is assumed much larger than $T$ such that the channel can be assumed constant with time over the whole grid. The signal received on the $(i,j)$'th grid location is given by $y_{ij}=\mathbf{x}^T_{ij} \mathbf{h}_j+n_{ij}$ at time slot $i$ and subcarrier $j$, where $\mathbf{x}_{ij} \in \mathbb{C}^N$ denotes the vector of downlink transmitted signal and $\mathbf{h}_m \in \mathbb{C}^N$ is the channel vector over the $j$'th subcarrier.\color{black}

Denoting downlink pilot signals transmitted by the BS over the $m^{th}$ subcarrier by $\mathbf{P}_m \in \mathbb{C}  ^{L \times N}$, where $L$ is the pilot length, the received signal at the user is given by
\small
\begin{align}\label{Model}
    \mathbf{y}_m=\mathbf{P}_m \mathbf{h}_m + \mathbf{n}_m,
\end{align}
\normalsize
where $\mathbf{n}_m \sim \mathcal{CN} (0,\sigma^2)$ is the complex additive Gaussian noise over the $m^{th}$ subcarrier, which is independent across subcarriers and time slots. A large pilot length $L$ is infeasible not only because it increases the training overhead and computational complexity for channel estimation, but also because $L$ should be much smaller than the channel coherence interval. Hence, in massive MIMO systems, where $N$ is excessively large, the pilot length $L$ is typically much smaller than the number of antennas $N$ and Eq. (\ref{Model}) is severely underdetermined. In this work we propose a NN architecture to jointly design the optimum pilot signals $\mathbf{P}_1, \hdots, \mathbf{P}_m \in \mathbb{C}  ^{L \times N}$ and to estimate the downlink channel matrix $\mathbf{H}$ from the signals received over all subcarriers, i.e., $\mathbf{Y}= [\mathbf{y}_1, \mathbf{y}_2, \hdots, \mathbf{y}_{M}]^T \in \mathbb{C}  ^{M \times L}$. 

Conventional channel estimation techniques are typically based on LMMSE estimation method, where the pilot signals are orthogonal discrete cosine transform (DFT) basis vectors or Zadoff-Chu sequences. The LMMSE channel estimate is given by $\mathbf{\hat{h}}_m^{MMSE}=\mathbb{E}[\mathbf{h}_m]+\mathbf{R}_{h_m y_m} \mathbf{R}_{y_m y_m}^{-1} (\mathbf{y}_m-\mathbb{E}[\mathbf{y}_m])$, where $\mathbb{E}[\mathbf{h}_m]$ and $\mathbb{E}[\mathbf{y}_m]$ are the corresponding expected values, and $\mathbf{R}_{h_m y_m}$ and $\mathbf{R}_{y_m y_m}$ are the covariance matrices over the $m$'th subcarrier. LMMSE is based on Gaussian channel assumption and requires the knowledge of the covariance matrices. In practical scenarios, the covariance matrices need to be empirically estimated over a dataset, which imposes additional computational load to the LMMSE technique. On the other hand, the simple choice of orthonormal DFT or Zadoff-Chu pilot signals regardless of the specific medium characteristics, e.g., scatterer positions, environment geometry, carrier frequencies, etc., is sub-optimal as will be discussed further in the next section. Our proposed approach utilizes a NN architecture for data-driven channel estimation and frequency-aware pilot design. The NN learns the required statistics over the dataset during training and leverages it for efficient channel estimation and pilot design; thereby reducing the pilot overhead. 


\begin{figure}[t!]
\centering
\includegraphics[scale=0.55]{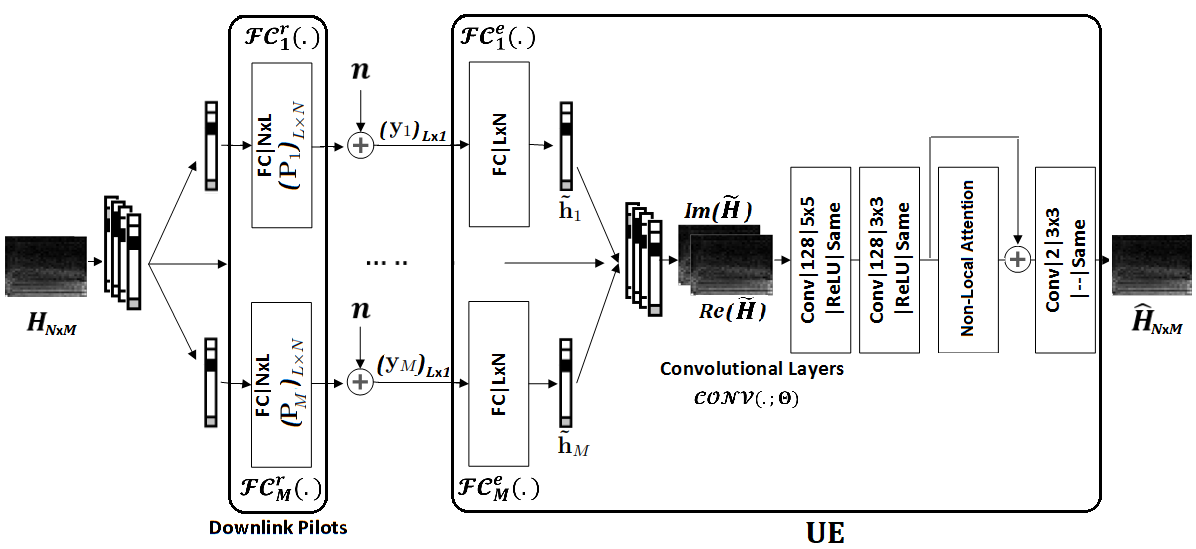}
\caption{\color{black}Block diagram of the proposed scheme.}
\label{BD}
\end{figure}

\section{NN-based pilot design and channel estimation}\label{sec2}

Fig. \ref{BD} depicts our proposed NN architecture for data-driven channel estimation and frequency-aware pilot design, where double channel inputs and outputs represent real and imaginary parts of the corresponding channel matrices. The network is composed of $M$ fully connected branches, one for each subcarrier, followed by 3 convolutional layers, and is trained in an end-to-end fashion to minimize the mean square error (MSE). Each branch is composed of 2 dense layers, a reduction and an expansion layer denoted by $\mathcal{FC}^r$ and $\mathcal{FC}^e$, respectively.

The input CSI matrix $\mathbf{H}$ is first divided into its subcarrier components $\mathbf{h}_1, \mathbf{h}_2, \hdots, \mathbf{h}_{M}$, where $\mathbf{h}_{m}$ is input to the $m$'th fully connected branch. The reduction layer in each branch models downlink pilot transmission over that subcarrier, which is followed by downlink additive Gaussian noise according to $\mathbf{y}_m=\mathcal{FC}^r_m(\mathbf{h}_{m}; \mathbf{P}_m)+ \mathbf{n}_m=\mathbf{P}_m \mathbf{h}_m + \mathbf{n}_m$. Hence, the weight parameters trained for $\mathcal{FC}^r_m$ correspond to the pilots to be transmitted over the $m$'th subcarrier, denoted by $\mathbf{P}^*_m$. Note that the dense layers $\mathcal{FC}^r_1, \hdots, \mathcal{FC}^r_M$ are used without bias or activations here, as they imitate the actual pilot transmission process, which is a simple matrix multiplication. A total power constraint on the pilot block is enforced by normalization, i.e. $\sum_{m=1}^M \|\mathbf{P}_m\|^2=1$. \color{black}Here, we implement complex-valued fully connected layers, i.e., we use 

\small
\begin{align}\label{REIM}
    \begin{pmatrix} \Re(\mathbf{y}_m) \\ \Im(\mathbf{y}_m) \end{pmatrix}=\begin{pmatrix} \Re(\mathbf{P}_m) & -\Im(\mathbf{P}_m) \\ \Im(\mathbf{P}_m) & \Re(\mathbf{P}_m) \end{pmatrix}\begin{pmatrix} \Re(\mathbf{h}_m) \\ \Im(\mathbf{h}_m) \end{pmatrix}+\begin{pmatrix} \Re(\mathbf{n}_m) \\ \Im(\mathbf{n}_m) \end{pmatrix},
\end{align}
\normalsize
where $\Re(\mathbf{P}_m)$ and $\Im(\mathbf{P}_m)$ represent real and imaginary parts of the pilots, respectively, and are trainable NN parameters. Hence, each fully connected layer has $2LN$ real-valued trainable parameters. \color{black}

The expansion layer in the $m$'th branch gives a coarse initial estimate of the channel at the $m$'th subcarrier according to $\mathbf{\Tilde{h}}_m=\mathcal{FC}^e_m(\mathbf{y}_{m}; \mathbf{Q}_m, \mathbf{b}_m)=\mathbf{Q}_m \mathbf{y}_m + \mathbf{b}_m$, \color{black} where the expansion layer is also implemented as a complex-valued fully connected layer, similarly to (\ref{REIM}).\color{black}  This initial estimate imitates the familiar LMMSE estimate $\mathbf{\hat{h}}_m^{MMSE}=\mathbb{E}[\mathbf{h}_m]+\mathbf{R}_{h_m y_m} \mathbf{R}_{y_m y_m}^{-1} (\mathbf{y}_m-\mathbb{E}[\mathbf{y}_m])$, which is the best linear estimator that minimizes $\mathbb{E}\| \mathbf{h}_m-\hat{\mathbf{h}}_m\|_2^2$, and the NN learns the linear weights $\mathbf{Q}_m$ and $\mathbf{b}_m$ to minimize the empirical average MSE given by $1/K \sum_{k=1}^K \| \mathbf{h}_m^{(k)}-\hat{\mathbf{h}}_m^{(k)}\|_2^2$, where $K$ is the size of the dataset.

We subsequently apply convolutional layers on these initial estimates to further improve the estimation accuracy by exploiting the inherent correlations within the channel matrix across subcarriers as well as antennas. This is motivated by our previous observations showing strong local correlations in MIMO channel matrices, which also explains the significant success of convolutional NNs in efficient MIMO CSI reduction and feedback in previous works \cite{yang2019deep, mashhadi2019cnnbased, yang2019deepJ}. To this end, we concatenate initial estimates from the dense branches to get $\mathbf{\Tilde{H}}=[ \mathbf{\Tilde{h}}_1, \hdots, \mathbf{\Tilde{h}}_M]$, which is then input to the convolutional layers, as shown in Fig. \ref{BD}. \color{black}The convolutional layers implement real-valued convolution kernels and treat the complex output of the expansion layers as a two-dimensional input by concatenating its real and imaginary parts along the channel dimension. \color{black}

In Fig. \ref{BD}, ``Conv$|128|$ $5\times5| $ReLU$|$Same'' represents a convolutional layer with 128 kernels of size $5\times5$ with rectified linear (ReLU) activation and ``Same'' padding technique. ``$|--|$'' means that the last layer has no activation. The final estimate is given by $\mathbf{\hat{H}}=\mathcal{CONV}(\mathbf{\Tilde{H}};\Theta)$, where $\mathcal{CONV}$ and $\Theta$ denote the convolutional layers and their corresponding set of parameters.

\color{black}The convolutional layers learn and exploit local correlations to improve the estimation accuracy. As we will see in section \ref{sec5}, local correlations account for a significant improvement in the reconstruction NMSE in comparison with simple LMMSE. However, we see that there still exists long range correlations in the channel matrix. In order to exploit long range correlations in an efficient manner, we have added a non-local attention module \cite{attention1, attention2}. We later show through simulations that the non-local attention module further improves the NMSE, specifically when fewer pilots are used and at lower downlink SNRs. 

\begin{figure}[t!]
\centering
\includegraphics[scale=0.8]{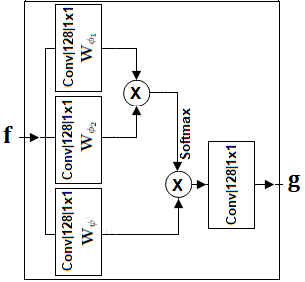}
\caption{\color{black}Block diagram of the non-local attention module.}
\label{attentionBD}
\end{figure}

Eq. (\ref{attention}) provides the general input-output relation for non-local attention:
\small
\begin{align}\label{attention}
    \mathbf{g}_i=\frac{1}{\pi(\mathbf{f})} \sum_{\forall j} \phi(\mathbf{f}_i, \mathbf{f}_j) \psi(\mathbf{f}_j),
\end{align}
\normalsize
in which $i$ is the spacetime index of an output position whose response is to be computed, $j$ is the
index that enumerates all possible positions, $\mathbf{f}$ is the input, and $\mathbf{g}$ is the output of the same size as $\mathbf{f}$. A pairwise function $\phi$ computes a scalar representing the relationship between $i$ and all $j$. The unary function $\psi$ computes a representation of the input signal at position $j$. The response is normalized by a factor $\pi(\mathbf{f})$. We tried various popular choices for $\pi(\cdot), \phi(\cdot)$  and $\psi(\cdot)$ functions (refer to \cite{attention2} for further details), and we found that the best performing ones in our problem are $\phi(\mathbf{f}_i, \mathbf{f}_j)=\exp{([\mathbf{W}_{\phi_1} \mathbf{f}_i]^T[\mathbf{W}_{\phi_2} \mathbf{f}_j])}$, $\psi(\mathbf{f}_j)=\mathbf{W}_{\psi} \mathbf{f}_j$, and $\pi(\mathbf{f})=\sum_{\forall j} \phi(\mathbf{f}_i, \mathbf{f}_j)$, where $\mathbf{W}$s are trainable weight matrices. Fig. \ref{attentionBD} provides the block diagram for our non-local attention module, where $1 \times 1$ convolutions implement $\mathbf{W}$ weight multiplications, and ``Softmax" activation implements the exponential function. Refer to \cite{attention2} for more details on attention modules. 
\color{black}  

Finally, the network is trained to minimize the end-to-end empirical MSE cost given by
\small
\begin{align}\label{MSE}
    \mathcal{C}(\mathbf{H};\mathbf{P}_1, \hdots, \mathbf{P}_M, \mathbf{Q}_1, \hdots, \mathbf{Q}_M, \Theta)&=\frac{1}{K} \sum_{k=1}^K MSE(\mathbf{H}^{(k)},\hat{\mathbf{H}}^{(k)})\\ \nonumber 
    & =\frac{1}{K} \sum_{k=1}^K \| \mathbf{H}^{(k)}-\mathcal{CONV}(\mathbf{\Tilde{H}}^{(k)};\Theta)\|_2^2,
\end{align}
\normalsize
where $H^(k)$ denotes the $k$'th sample channel in the training dataset and $K$ is the dataset size. Minimizing (\ref{MSE}) end-to-end, we obtain all the network parameters including the pilot signals $\mathbf{P}_1, \hdots, \mathbf{P}_M$.


\textbf{Remark 1:} In comparison with the simple LMMSE estimator, our proposed NN-based approach is data-driven; in the sense that, it does not make any prior assumption on the channel statistics, but rather learns it from the dataset during training. Although LMMSE is optimum when channel and noise distributions are jointly Gaussian, the general MMSE estimate $\mathbf{\hat{H}}=\mathbb{E}[\mathbf{H}|\mathbf{Y}]$ has no closed-form expression for arbitrary channel distributions. Our proposed NN-based channel estimator is trained to minimize the empirical MSE, and hence, gives a tractable approximation of the general MMSE estimate regardless of the channel distribution. Moreover, our proposed NN architecture uses convolutional layers to exploit inter-frequency and inter-antenna correlations to improve the estimation performance. We later show through simulations that the proposed NN-based approach outperforms LMMSE channel estimation. The improvement is more significant for shorter pilot lengths and lower SNR values. 

\textbf{Remark 2:} There has been a prior line of research to analytically design optimum pilot signals for downlink FDD MIMO systems \cite{STP1, STP2, STP3}. When the channel distribution is assumed to be Gaussian, the resulting MSE of the LMMSE estimator has a closed-form expression. The authors in \cite{STP2} use steepest descent optimization to design pilots that minimize the resulting MSE. This approach again relies on the Gaussian channel assumption. If this assumption is slightly generalized to a Gaussian mixture distribution, then closed-form expressions are available only for upper and lower bounds on the MSE \cite{STP3}. The authors in \cite{STP3} use steepest descent to design pilots that maximize the mutual information between the received noisy pilots and the channel, i.e., $I(\mathbf{Y};\mathbf{H})$. Our NN approach, however, jointly designs channel estimator and pilot signals to minimize the end-to-end estimation error while avoiding computational difficulties.

\textbf{Remark 3:} The authors in \cite{Navid, DLest1} utilize a CNN-based architecture for MIMO channel estimation, but they use simple FFT pilot signals. In this paper, we add the fully connected reduction layers to the CNN architecture to jointly optimize pilots and estimate the channel. We show later in the ablation study that the designed pilot signals significantly improve the performance. They also use the CNN structure to improve the NMSE of an initial LS estimate. We observed in simulations that replacing the initial LS estimate with the fully connected expansion layers lets the network to automatically learn the initial estimate and improves the end-to-end reconstruction MSE. Although the approach proposed in \cite{Navid} and \cite{DLest1} outperforms non-ideal LMMSE (where the covariance matrices are estimated from a coarse initial LS estimate at the user), it still performs worse than the ideal LMMSE. We later show through simulations that our proposed NN-based approach outperforms ideal LMMSE (where the covariance matrices are estimated over the dataset). The improvement is more significant for shorter pilot lengths and lower SNR values.

\textbf{Remark 4:} The authors in \cite{EstPil} use a similar NN structure, but they utilize the same pilot signals over all the subcarriers. However, we observe different channel statistics over different subcarriers, which suggests that using the same pilots over all the subcarriers is sub-optimal. In this work, we use a frequency-aware pilot design approach, which utilizes different fully connected branches over different subcarriers to design pilots. We later show through simulations that our frequency-aware approach improves the MSE by a considerable margin. It also allows us to non-uniformly allocate pilots to different subcarriers by our proposed NN pruning technique, presented in the next section, which further reduces the MSE with the same pilot overhead.

\section{Pilot Allocation by NN pruning}\label{sec4}

In FDD operation mode of massive MIMO systems, downlink pilot transmission consumes a significant amount of radio resources, which in turn results in a significant loss of spectral efficiency. While the NN architecture proposed in the previous section significantly improves the end-to-end channel estimation performance by exploiting inherent correlations in the channel matrix using convolutional layers and designing pilot signals in a frequency-aware manner using the fully connected branches, our goal in this section is to further reduce the pilot overhead. To achieve this, we propose an efficient pilot allocation technique by pruning the least significant neurons from the fully connected layers in the proposed NN architecture. The reduction layer in the $m$'th fully connected branch consists of $L$ neurons each of which corresponds to a single pilot transmission according to $\mathbf{\Tilde{y}}_m=\mathcal{FC}^r_m(\mathbf{h}_{m}; \mathbf{P}_m)=\mathbf{P}_m \mathbf{h}_m$, and occupies one time-frequency resource over the downlink channel. We denote the pilot matrix by $\mathbf{\Tilde{Y}}=[\mathbf{\Tilde{y}}_1, \mathbf{\Tilde{y}}_2, \hdots, \mathbf{\Tilde{y}}_M]$, which represents a total of $L \times M$ time-frequency resources allocated to downlink pilot transmission during each coherence interval of the channel. Our idea is to gradually prune least sigificant neurons from the fully connected reduction layers $\mathcal{FC}^r_1, \hdots, \mathcal{FC}^r_M$ during training to reduce the pilot overhead by saving the corresponding time-frequency resources for data transmission, while causing the least possible degradation to the reconstruction MSE. This approach enables non-uniform pilot allocation and the transmission of fewer number of pilots over subcarriers that can be satisfactorily reconstructed by the subsequent convolutional layers utilizing inter-frequency correlations.

\color{black}
Consequently, we formulate the design and optimization of the pilots as a NN pruning problem, where the goal is to simplify a large NN by pruning some weights/neurons from it without significantly degrading the performance \cite{Pruning1, Pruning2}. We highlight here that pruning of NNs is typically employed to reduce the computation and memory complexity of NNs, where as in our case, pruning the fully-connected reduction layers translates into reducing the pilot signals. Different works have heuristically proposed various saliency metrics to decide which connections/neurons can be pruned from a large NN without significantly degrading its performance. Saliency metrics heuristically approximate the relative importance of different weights or sets of weights in the NN. One of the most widespread used metrics is the $l_1$-norm of the weights \cite{Prun1}, which leads to the magnitude-based pruning scheme \cite{Pruning1}. This is motivated by the heuristic that the larger the weight value, the more likely the weight is to be contributing to the result. Other saliency metrics include average ratio of zero activations \cite{Prun2}, the Fisher information \cite{Prun3}, and 1st order Taylor expansions \cite{Prun4}, or a composition of various metrics \cite{Prun5}.
\color{black}

Inspired by \cite{Pruning1}, we take an $l_1$-regularized magnitude-based pruning approach by introducing two modifications to the training process. Note that the $(i,j)$'th pilot signal is given by

\small
\begin{align}\label{mask}
    [\mathbf{\Tilde{Y}}]_{ij}=[\mathbf{\Tilde{y}}_j]_i=[\mathcal{FC}^r_j(\mathbf{h}_{j}; \mathbf{P}_j)]_i=[\mathbf{P}_j \mathbf{h}_j]_i=\sum_{n=1}^N [\mathbf{P}_j]_{in}[\mathbf{h}_j]_n,
\end{align}
\normalsize
where $[.]_{ij}$ represents the $(i,j)$'th element of a matrix. We define matrix $\mathbf{\Phi}$, where $[\mathbf{\Phi}]_{ij}=\sum_{n=1}^N \|[\mathbf{P}_j]_{in}\|^2$. Our idea is to prune neurons with the smallest sum squared weight connections (i.e., the smallest $[\Phi]_{ij}$) from the network. With this in mind, we make the following modifications to the training process:

\textit{Sparsity promoting regularization:}
We add a second term to our cost function to push $[\mathbf{\Phi}]_{ij}$ elements towards zero. In particular, we add the $l_1$-norm of the $\mathbf{\Phi}$ matrix to the cost function with a regularization parameter $\lambda$, which controls the trade-off between sparsity and the MSE according to 
\small
\begin{align}\label{pruning}
    \mathcal{C}_{prun}(\mathbf{H};\mathbf{P}_1, \hdots, \mathbf{P}_M, \mathbf{Q}_1, \hdots, \mathbf{Q}_M, \Theta) & =\frac{1}{K} \sum_{k=1}^K MSE(\mathbf{H}^{(k)},\hat{\mathbf{H}}^{(k)}) + \lambda \|\mathbf{\Phi}\|_1 \\ \nonumber
    &=\frac{1}{K} \sum_{k=1}^K \| \mathbf{H}^{(k)}-\hat{\mathbf{H}}^{(k)}\|_2^2 + \lambda \sum_i\sum_j |\sum_{n=1}^N \|[\mathbf{P}_j]_{in}\|^2|,
\end{align}
\normalsize
where $\lambda$ is determined through numerical search until a desired sparsity level is achieved. Denoting our target sparsity level by $S$, which represents the percentage of the pruned neurons, a larger $\lambda$ is required for larger $S$ values.

\textit{Magnitude-based pruning:}
Define a pilot allocation mask $\mathbf{M} \in [0,1]^{L \times M}$, in which the zeros represent the pruned pilots. Pruning is performed by element-wise multiplication of the pruning mask with the received pilot matrix during each optimization step, i.e., $\mathbf{Y}=\mathbf{M} \odot \mathbf{\Tilde{Y}}+\mathbf{N}$. We initialize $\mathbf{M}$ to an all-one state and gradually update it during training according to a pruning schedule to achieve the final allocation mask $\mathbf{M^*}$. Our target sparsity $S$ represents the ratio of zeros in the final mask $\mathbf{M}^*$ over its size, $L \times M$. As our regularization term gradually pushes more and more of the $[\mathbf{\Phi}]_{ij}$ values towards zero along the training steps, $[\mathbf{M}]_{ij}$ values corresponding to the smallest $[\mathbf{\Phi}]_{ij}$'s can be zeroed out. We use a linear pruning schedule where we apply 10 balanced updates on $\mathbf{M}$ during training, each of which prunes another $S/10$ of the pilots. It is important to train for a sufficient number of steps after each update of the pruning mask to allow the network to converge to the new optimum before the next pruning update.

\color{black}
Although the choice of the saliency metric has been mostly heuristic in the literature, we do have some intuition to support our magnitude-based approach for the specific channel estimation task. Utilizing Eq. (\ref{mask}), the pilot power received on the $(i,j)$'th resource grid is bounded by
\small
\begin{align}\label{mask}
    \|[\mathbf{\Tilde{Y}}]_{ij}\|^2=\|\sum_{n=1}^N [\mathbf{P}_j]_{in}[\mathbf{h}_j]_n\|^2 \leq (\sum_{n=1}^N \|[\mathbf{P}_j]_{in}\|^2)(\sum_{n=1}^N \|[\mathbf{h}_j]_{n}\|^2)= [\Phi]_{ij} (\sum_{n=1}^N \|[\mathbf{h}_j]_{n}\|^2).\nonumber
\end{align}
\normalsize
Denoting the received pilot SNR on the $(i,j)$'th resource grid by $\rho_{ij}$, we get $\rho_{ij}=\|[\mathbf{\Tilde{Y}}]_{ij}\|^2/\sigma_n^2 \leq [\Phi]_{ij} (\sum_{n=1}^N \|[\mathbf{h}_j]_{n}\|^2/\sigma_n^2)$. Hence, the pilots located on grid locations with the smallest sum squared weight connections (i.e. the smallest $[\Phi]_{ij}$) are received at  lower SNRs and hence contribute the least to the NMSE performance achieved by the NN.
\color{black}

\begin{table}[!t] 
\centering 
\caption{Comparison of the NMSE (dB) values achieved by the proposed NN-based channel estimation scheme and the LMMSE technique for the indoor scenario ($N=32, M=256$).}\label{T11}
\resizebox{16cm}{!}{
\begin{tabular}{c|c|c|c|c|c|c|c}
\multicolumn{2}{c|}{ }   & $\mathrm{SNR}=-5 \mathrm{dB}$  & $\mathrm{SNR}=0 \mathrm{dB}$   & $\mathrm{SNR}=5 \mathrm{dB}$   & $\mathrm{SNR}=10 \mathrm{dB}$  & $\mathrm{SNR}=15 \mathrm{dB}$  & $\mathrm{SNR}=20 \mathrm{dB}$  \\ \hline
\multirow{2}{*}{$L=16$} & NN+Attention & -14.32 & -17.61 & -20.58 & -22.88 & -23.65 & -23.93 \\ \cline{2-8}
                        & NN & -13.40 & -16.87 & -19.95 & -22.46 & -23.40 & -23.75 \\ \cline{2-8}
                        & FC Subnet & -8.08 & -11.39 & -15.00 & -18.21 & -21.45 & -23.03 \\ \cline{2-8}
                      & LMMSE & -8.10 & -11.47 & -15.00 & -18.27 & -21.49 & -23.06 \\ \hline
\multirow{2}{*}{$L=12$} & NN+Attention & -11.33  & -13.61 & -14.83 & -15.35 & -16.22 & -16.79 \\ \cline{2-8} 
                        & NN & -8.39  & -11.53 & -13.35 & -14.41 & -15.52 & -16.40 \\ \cline{2-8} 
                        & FC Subnet & -4.75  & -8.00 & -11.07 & -13.10 & -13.94 & -14.80 \\ \cline{2-8}
                      & LMMSE & -4.82  & -8.08 & -11.17 & -13.15 & -13.96 & -14.82 \\ \hline
\multirow{2}{*}{$L=8$}  & NN+Attention & -13.23  & -13.96 & -14.41 & -14.66 & -14.82 & -15.04 \\ \cline{2-8}
                        & NN & -7.99  & -11.03 & -12.10 & -12.87 & -13.55 & -14.05 \\ \cline{2-8}
                        & FC Subnet & -3.75  & -6.98  & -8.95 & -10.11 & -11.09 & -11.80 \\ \cline{2-8}
                      & LMMSE & -3.87  & -7.05  & -9.01 & -10.15 & -11.11 & -11.82 \\ \hline
\end{tabular}}
\end{table}

\section{Simulation Results}\label{sec5}

\color{black}We have generated the datasets for training and testing using the COST 2100 channel model \cite{COST2100}, which is a geometry-based stochastic channel model that reproduce the stochastic properties of MIMO channels over time, frequency and space. COST 2100 is a cluster-level model, and generates channel realizations that hold statistical consistency of the large scale channel parameters. \color{black}We have considered an indoor picocellular scenario at 5.3 GHz and an outdoor rural scenario at 330 MHz band. The BS is equipped with a ULA of dipole antennas positioned at the center of a $20\mathrm{m} \times 20\mathrm{m}$ and $400\mathrm{m} \times 400\mathrm{m}$ square area for the indoor and outdoor scenarios, respectively. Note that we have presented the results for the outdoor scenario in Subsection V.B, but as the simulations revealed very similar results and conclusions for both the indoor and outdoor scenarios, we have included the results only for the indoor scenario in the subsequent subsections to avoid tedious discussions of similar results.

\color{black}The user is placed within the square area uniformly at random. All other parameters follow the default settings in \cite{COST2100}. Train and test datasets include 80000 and 20000 channel realizations, respectively, with $M=256$ and $N=32$. We train all NNs up to 110000 steps with a batch size of 100 utilizing the Adam optimizer \cite{kingma2014adam}. \color{black}We use the normalized MSE (NMSE) as the performance measure, defined by
\begin{equation}
\mathrm{NMSE} \triangleq \frac{ \mathbb{E} \{\|\mathbf{H}-\mathbf{\hat{H}}\|_2^2\}}{ \mathbb{E} \{\|\mathbf{H}\|_2^2 \}}.  
\end{equation}

\begin{figure}[t!]
\centering
\includegraphics[scale=0.55]{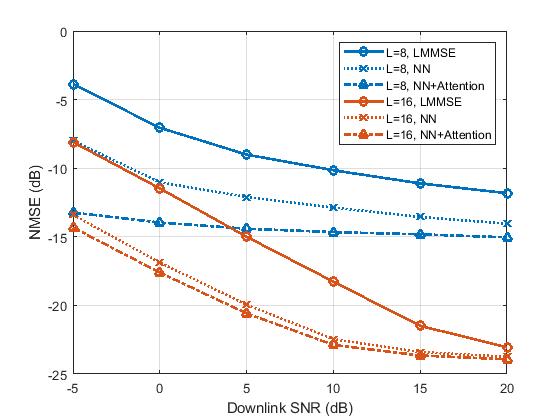}
\caption{NMSE (dB) vs. downlink SNR for the indoor scenario, $N=32, M=256$.}
\label{NNvsLMMSE}
\end{figure}

\subsection{NN vs. LMMSE performance}

Table \ref{T11} and Fig. \ref{NNvsLMMSE} compare the NMSE values achieved by the proposed NN-based channel estimation technique with those achieved by LMMSE for the indoor scenario at different pilot length and SNR values. We note that we have used the pilot signals optimized by the fully connected reduction layers for the LMMSE approach here. For LMMSE, we first estimate the channel covariance matrix over the training dataset utilizing the maximum likelihood (ML) covariance estimation for Gaussian vectors \cite{Kay}, and then use it to get the LMMSE channel estimate. Note that we have assumed that the user has perfect knowledge of these estimated covariance matrices. In practice, the user may need to estimate the covariance matrix from a coarse initial LS estimate of the channel \cite{DLest1}, which degrades the LMMSE performance. Hence, the curves in Fig. \ref{NNvsLMMSE} provide a lower bound on the performance of LMMSE as we have used covariance matrices estimated over the original training dataset. \color{black} Table \ref{T11} also presents the contribution of each subnetwork of the proposed architecture towards the final NMSE value. In this table, we report the NMSE values for NN-based channel estimation with and without the attention module, denoted by ``NN+Attention" and ``NN", respectively. ``FC subnet" corresponds to the NMSE obtained when only the fully connected layers are used for NMSE evaluations. 
\color{black}

%

\color{black}According to Fig. \ref{NNvsLMMSE}, the NN-based approach outperforms LMMSE for all SNR and pilot lengths. The proposed NN-based approach even without the attention module outperforms LMMSE, and the attention module further improves the performance specifically at low SNR regimes. For example, at a downlink SNR=-5dB, the ``NN+Attention" scheme improves the reconstruction NMSE by $9.36 \mathrm{dB}$ , $6.51 \mathrm{dB}$, and $6.22 \mathrm{dB}$ in comparison with LMMSE for $L=8, 12$ and $16$, respectively. For a target NMSE of $\sim -15$dB at a downlink SNR of 5dB, our proposed NN+Attention approach reduces the required pilot length from 16 to 8 in comparison with LMMSE, resulting in a 50\% reduction in the pilot overhead.

Finally, comparing the ``LMMSE" and ``FC Subnet" results in Table \ref{T11} shows that the FC subnetwork achieves almost the same NMSE performance as the LMMSE estimate. We have also observed that the parameters learned by the FC layers, are very close to the LMMSE estimator, i.e., $\mathbf{Q}_m \sim \mathbf{R}_{h_m y_m} \mathbf{R}_{y_m y_m}^{-1}$ and $\mathbf{b}_m \sim \mathbb{E}[\mathbf{h}_m]-\mathbf{R}_{h_m y_m} \mathbf{R}_{y_m y_m}^{-1}\mathbb{E}[\mathbf{y}_m]$, which shows that the initial estimate at the output of the FC expansion layers (i.e., $\mathbf{\Tilde{H}}$) imitates the LMMSE estimate.
\color{black}
 
 \begin{table}[!t] 
\centering 
\caption{Comparison of the NMSE (dB) values achieved by the proposed NN-based channel estimation scheme and the LMMSE technique for the outdoor scenario ($N=32, M=256$).}\label{T12}
\resizebox{16cm}{!}{
\begin{tabular}{c|c|c|c|c|c|c|c}
\multicolumn{2}{c|}{ }   & $\mathrm{SNR}=-5 \mathrm{dB}$  & $\mathrm{SNR}=0 \mathrm{dB}$   & $\mathrm{SNR}=5 \mathrm{dB}$   & $\mathrm{SNR}=10 \mathrm{dB}$  & $\mathrm{SNR}=15 \mathrm{dB}$  & $\mathrm{SNR}=20 \mathrm{dB}$  \\ \hline
\multirow{2}{*}{$L=16$} & NN+Attention & -16.84 & -18.82 & -20.37 & -20.83 & -21.34 & -21.55 \\ \cline{2-8} 
                        & NN & -15.31 & -17.89 & -19.68 & -20.36 & -20.84 & -21.14 \\ \cline{2-8} 
                      & LMMSE & -7.82 & -11.92 & -15.75 & -18.91 & -19.67 & -20.47 \\ \hline
\multirow{2}{*}{$L=12$} & NN+Attention & -10.04  & -11.05 & -11.83 & -12.64 & -13.25 & -13.82 \\ \cline{2-8}
                        & NN & -8.37  & -9.99 & -11.41 & -12.25 & -12.98 & -13.61 \\ \cline{2-8}
                      & LMMSE & -3.76  & -7.20 & -10.33 & -11.38 & -12.15 & -12.89 \\ \hline
\multirow{2}{*}{$L=8$}  & NN+Attention & -7.37  & -7.76 & -8.06 & -8.37 & -8.57 & -8.76 \\ \cline{2-8} 
                        & NN & -4.91  & -6.29 & -7.34 & -7.68 & -7.92 & -8.07 \\ \cline{2-8} 
                      & LMMSE & -1.88  & -3.51  & -4.96 & -6.39 & -6.67 & -6.89 \\ \hline
\end{tabular}}
\end{table}

\color{black}
\subsection{Impact of the scattering environment}
\begin{figure}[t!]
\centering
\includegraphics[scale=0.55]{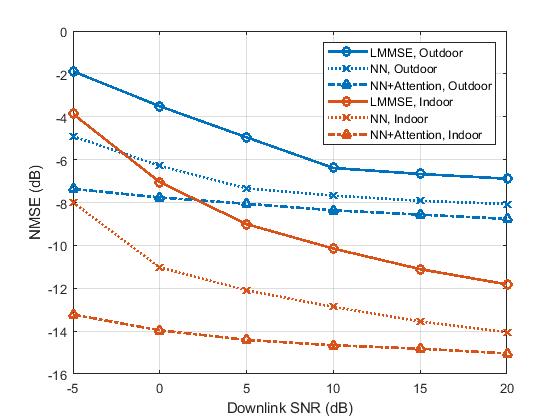}
\caption{NMSE (dB) comparisons for the indoor vs. outdoor scenarios, $L=8, N=32, M=256$.}
\label{NNvsLMMSEOutdoor}
\end{figure}

Table \ref{T12} provides the reconstruction NMSEs for the outdoor scenario at different pilot length and SNR values. Note that we have used the pilot signals optimized by the fully connected reduction layers for the LMMSE approach here. Fig. \ref{NNvsLMMSEOutdoor} compares the NMSE values for the indoor and outdoor scenarios when $L=8, N=32, M=256$. This figure shows a more significant performance improvement achieved by the NN-based approach for the outdoor scenario in comparison with LMMSE. According to Tables \ref{T11} and \ref{T12}, the ``NN+Attention" approach improves the reconstruction NMSE by $5.49 \mathrm{dB}$ and $9.36\mathrm{dB}$ in comparison with LMMSE for the outdoor and indoor scenarios, respectively at $L=8, N=32, M=256$. 

\color{black}

\subsection{Impact of channel SNR}
\begin{figure}[t!]
\centering
\includegraphics[scale=0.55]{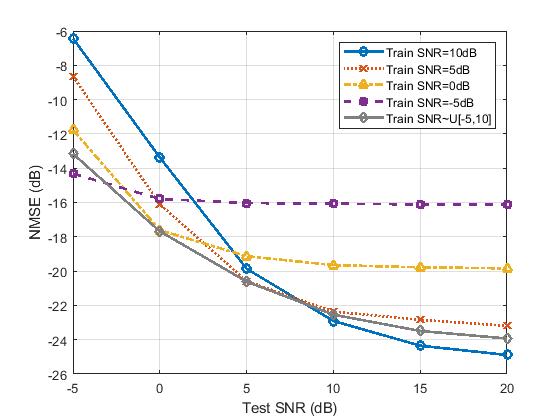}
\caption{NMSE (dB) vs. test SNR for networks trained at different SNR values with $L=16, N=32, M=256$.}
\label{SNR}
\end{figure}

Fig. \ref{SNR} studies the performance of the NN-based+Attention scheme when there is a mismatch between the training and test SNRs. This figure plots the NMSE versus test SNR curves for networks trained with each specific SNR value. We see that the NMSE degrades when the test SNR falls below the SNR value used for training. On the other hand, for a network trained with a lower SNR value, the performance saturates when test SNR improves above the training SNR. The grey curve represents the performance of a NN trained over a dataset with sample SNRs picked uniformly at random from the interval $[-5,10]$dB. As observed in this figure, such a NN performs satisfactorily over the whole SNR range.

\subsection{Frequency-aware pilot design}
Unlike in \cite{EstPil}, which optimizes the same pilots over all subcarriers, our proposed NN architecture utilizes $M$ parallel fully connected branches to design different pilots over the subcarriers. This helps to improve the NMSE performance by allowing the NN to learn and exploit specific subcarrier statistics and structures in a frequency-aware manner. Table \ref{T3} provides a comparison between our proposed frequency-aware pilot design scheme with the approach proposed in \cite{EstPil}, where the same pilots are used over different subcarriers. The acronyms ``SP" and ``DP" in this table represent ``same pilots" and ``different pilots" approaches, respectively. 

\begin{figure}[t!]
\centering
\includegraphics[scale=.5]{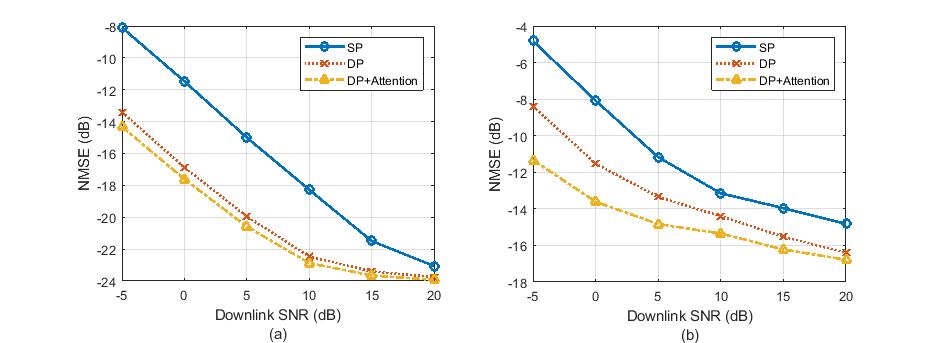}
\caption{Comparison between the NMSE (dB) values achieved using same pilots (SP) over all subcarriers \cite{EstPil} and the proposed frequency-aware pilot design scheme (DP) for $N=32, M=256$, (a) $L=16$, (b) $L=12$.}
\label{SPDP}
\end{figure}

\color{black}According to Table \ref{T3}, the proposed frequency-aware approach improves the performance significantly even without the attention module, e.g., for a target NMSE of $-12.9$dB at a downlink SNR of 10dB, the ``DP" approach reduces the required pilot length from 12 to 8 in comparison with ``SP", which is a 33.3\% reduction in the pilot overhead. The improvements are even more significant when the attention module is introduced to the proposed frequency-aware approach, e.g. for  a downlink SNR of $-5$dB, the ``DP+attention" approach reduces the required pilot length from 16 to 8 in comparison with ``SP" and the reconstruction NMSE by 1.39dB. Fig. \ref{SPDP} compares the NMSE curves versus downlink SNR for ``SP", ``DP" and ``DP+Attention" schemes for (a) $L=16$ and (b) $L=12$. This figure also shows a considerable NMSE improvement by the proposed ``DP+Attention" scheme in comparison with the ``SP" approach proposed in \cite{EstPil}.\color{black}


\begin{table}[!t] 
\centering 
\caption{Comparison of the NMSE (dB) achieved using same pilots (SP) over all subcarriers \cite{EstPil} and the proposed frequency-aware pilot design scheme (DP) for the indoor scenario at $N=32, M=256$.}\label{T3}
\resizebox{16cm}{!}{
\begin{tabular}{c|c|c|c|c|c|c|c}
\multicolumn{2}{c|}{ }   & $\mathrm{SNR}=-5 \mathrm{dB}$  & $\mathrm{SNR}=0 \mathrm{dB}$   & $\mathrm{SNR}=5 \mathrm{dB}$   & $\mathrm{SNR}=10 \mathrm{dB}$  & $\mathrm{SNR}=15 \mathrm{dB}$  & $\mathrm{SNR}=20 \mathrm{dB}$  \\ \hline
\multirow{2}{*}{$L=16$} & DP+Attention & -14.32 & -17.61 & -20.58 & -22.88 & -23.65 & -23.93 \\ \cline{2-8}
                        & DP & -13.40 & -16.87 & -19.95 & -22.46 & -23.40 & -23.75 \\ \cline{2-8}
                      & SP & -11.84 & -15.07 & -18.08 & -20.58 & -22.34 & -23.25 \\ \hline
\multirow{2}{*}{$L=12$} & DP+Attention & -11.33  & -13.61 & -14.83 & -15.35 & -16.22 & -16.79 \\ \cline{2-8}
                        & DP & -8.39  & -11.53 & -13.35 & -14.41 & -15.52 & -16.40 \\ \cline{2-8}
                      & SP & -7.92  & -10.92 & -11.74 & -12.91 & -13.30 & -13.94 \\ \hline
\multirow{2}{*}{$L=8$}  & DP+Attention & -13.23  & -13.96 & -14.41 & -14.66 & -14.82 & -15.04 \\ \cline{2-8}
                        & DP & -7.99  & -11.03 & -12.10 & -12.87 & -13.55 & -14.05 \\ \cline{2-8}
                      & SP & -7.74  & -9.38  & -10.69 & -12.12 & -12.99 & -13.42 \\ \hline
\end{tabular}}
\end{table}

\subsection{Ablation study}
\begin{figure}[t!]
\centering
\includegraphics[scale=0.55]{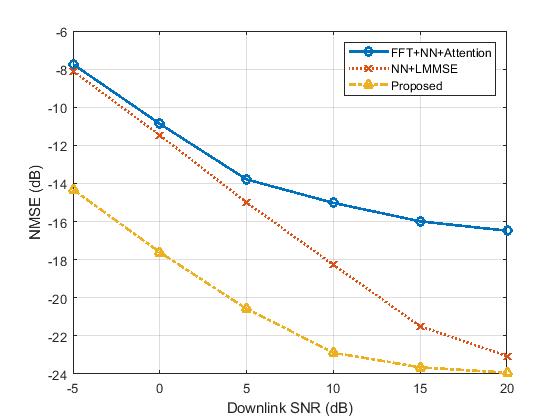}
\caption{Ablation results of NMSE (dB) for $L=16$, $N=32, M=256$.}
\label{Ablation}
\end{figure}

To study the improvements by each component of the proposed NN architecture, we provide ablation results in Table \ref{T4} and Fig. \ref{Ablation}. We compare 4 different schemes as below:
\begin{itemize}
    \item Our proposed scheme, in which both the pilot design and channel estimation are handled by the NN including the attention module. This scheme is denoted by ``NN+NN+Attention" in the table.
    \item A scheme, which uses LMMSE for channel estimation while the pilots are designed using a NN. In this case we take the received pilot vectors  $\mathbf{y}_m=\mathbf{P}_m^* \mathbf{h}_m + \mathbf{n}_m$ from the fully connected branches and apply ideal LMMSE (utilizing covariances estimated over the dataset) on it, where $\mathbf{P}_m^*$ denotes the pilots optimized by the NN. This scheme is denoted by ``NN+LMMSE" in the table.
    \item A scheme that employs the ``NN+Attention" for channel estimation, but simple FFT pilots are used as proposed in \cite{Navid} and \cite{DLest1}. Note that, in this scheme, we train the NN+Attention module replacing $\mathbf{P}_m$'s with a submatrix of DFT. This scheme is denoted by ``FFT+NN+Attention" in the table.
    \item The conventional scheme using FFT pilots and LMMSE estimation. This scheme is denoted by ``FFT+LMMSE" in the table.
\end{itemize}

According to Table \ref{T4} and Fig. \ref{Ablation}, the proposed fully NN-based scheme outperforms significantly all the three schemes with the NN-based pilot design leading to a remarkable improvement in the reconstruction NMSE. The conventional FFT+LMMSE scheme performs poorly at all channel SNRs as channel estimation is severely underdetermind for $N=32$ and $L=16$.

\color{black}
\subsection{Comparison with extended LMMSE}
We showed in the previous subsections that the proposed NN-based channel estimator outperforms the conventional LMMSE estimate by exploiting the inter-frequency correlations in the MIMO channel matrix. In particular, convolutional layers learn and exploit local correlations to improve the estimation accuracy. Local correlations account for a significant performance improvement in the reconstruction NMSE in comparison with simple LMMSE. However, we see that there still exist long range correlations in the channel matrix that are exploited by the proposed attention modules to further improve the reconstruction accuracy. 

\begin{table}[!t] 
\centering 
\caption{Ablation results of NMSE (dB) for $L=16$, $N=32, M=256$.}\label{T4}
\resizebox{16cm}{!}{
\begin{tabular}{c|c|c|c|c|c|c}
 { } & $\mathrm{SNR}=-5 \mathrm{dB}$  & $\mathrm{SNR}=0 \mathrm{dB}$   & $\mathrm{SNR}=5 \mathrm{dB}$   & $\mathrm{SNR}=10 \mathrm{dB}$  & $\mathrm{SNR}=15 \mathrm{dB}$  & $\mathrm{SNR}=20 \mathrm{dB}$ \\
\hline 
 Proposed (NN+NN+Attention) & -14.32  & -17.61 &  -20.58 & -22.88 & -23.65 & -23.93\\
\cline{1-7} 
NN+LMMSE & -8.10 & -11.47 &  -15.00 & -18.27 & -21.49 & -23.06\\
\cline{1-7}
 FFT+NN+Attention & -7.74 & -10.87 & -13.79 & -15.02 & -15.99 & -16.48\\
 \cline{1-7}
 FFT+LMMSE & -0.86 & -0.92 & -1.19 & -1.77 & -2.77 & -4.17\\
\hline
\end{tabular}}
\end{table}

Note that one can similarly use an extended LMMSE approach which exploits all the inter-frequency and inter-antenna correlations in the channel matrix assuming a Gaussian channel model. This approach is  computationally prohibitive as it requires an accurate estimate of a large covariance matrix of size $NM \times NM$ and becomes infeasible in wideband massive MIMO systems where both $N$ and $M$ are significantly larger. Hence, in practical scenarios, the subcarrier-wise LMMSE estimate, which exploits inter-antenna correlations in each subcarrier is used, and hence, this subcarrier-wise LMMSE estimate has been used for comparisons in previous subsections as well as previous works \cite{Navid, DLest1}.  

\begin{table}[!t] 
\centering 
\caption{\color{black}Parameter complexity of various channel estimators.}\label{Complexity}
\resizebox{9cm}{!}{%
\color{black}\begin{tabular}{c|c}
   &  Number of trainable parameters  \\
\hline 
 DP+Attention &  $2LNM+C_{1}+C_{2}$\\\cline{1-2}
 
 DP &  $2LNM+C_{1}$\\\cline{1-2}
 
 SP &  $2LN+C_{1}$\\\cline{1-2}
 
 LMMSE &  $2N^2  M+2NM+1$ \\\cline{1-2}
 
 Extended LMMSE &  $2N^2  M^2+2NM+1$\\\cline{1-2}
\hline
\end{tabular}\color{black}}
\end{table}

However, to understand the full potential of the proposed datat-driven approach, in this subsection, we compare the performance of our proposed NN-based approach with the extended LMMSE bound. We are specifically interested in understanding the significance of long range correlations, and how well our proposed NN+Attention approach can learn and exploit these correlations. For the extended LMMSE we estimate the required covariance matrix over the training dataset utilizing the optimum estimator \cite{Kay}. Hence, the number of real-valued trainable parameters for the extended LMMSE equals $2N^2M^2+2NM+1$, where $2N^2M^2$ of the parameters account for the covariance matrix elements, $2NM$ account for the expected channel values and $1$ is for the noise variance. 

Table \ref{Complexity} provides the number of trainable parameters for several channel estimators considered in this paper. Note that the number of parameters for the convolutional layers and the attention module is independent of the dimensions of the channel matrix. This is very favorable as these modules can be used in wide-band massive MIMO systems without increasing the parameter complexity, while the model complexity increases with $N^2$ and $M^2$ for extended LMMSE and becomes infeasible in wideband massive MIMO scenarios. The number of trainable parameters for the convolutional layers (denoted by $C_1$) and the attention module (denoted by $C_2$) for our proposed architectures in Fig. \ref{BD} and \ref{attentionBD} are given by:
\small
\begin{align}\label{Cs}
   C_1 & = 128(5\times5+1)+128(3\times3+1)+2(3\times3+1)=4628, \\\nonumber
   C_2 & = 4\times128(128\times1\times1+1)=66048.
\end{align}
\normalsize

Table \ref{extendedperf} presents the reconstruction NMSE for various channel estimator when trained on datasets of various sizes. Note that we originally trained all NNs on a dataset of size $|\mathcal{D}|=80K$, which includes 80K random realizations of the MIMO-OFDM channel matrices. In this table, we evaluate the performance of various channel estimators when less data is available for training. According to Table \ref{extendedperf}, although the extended LMMSE slightly outperforms the NN-based approach when a large dataset is available to estimate the covariance matrix, its performance degrades significantly when less data is available for training. This is due to the fact that the extended LMMSE approach requires estimating a large covariance matrix with 134M trainable parameters and when less data is available for training, estimates of these parameters become less accurate leading to a significant performance loss. The performance of all the other approaches remains almost the same for smaller dataset sizes.

\begin{table}[!t] 
\centering 
\caption{\color{black}NMSE comparisons at $\mathrm{SNR}=-5 \mathrm{dB}$ and $L=16, N=32, M=256$.}\label{extendedperf}
\resizebox{13cm}{!}{%
\color{black}\begin{tabular}{c|c|c|c|c|c}
   & DP+Attention & DP & SP & LMMSE & Extended LMMSE \\
\hline 
 $|\mathcal{D}|=10K$ &  -14.28 & -13.41 &  -11.85 & -8.10 & -12.14\\\cline{1-6}
 $|\mathcal{D}|=20K$ &  -14.26 & -13.38 &  -11.80 & -8.10 & -13.45\\\cline{1-6}
 $|\mathcal{D}|=40K$ &  -14.30 & -13.43 &  -11.88 & -8.08 & -13.94\\\cline{1-6}
 $|\mathcal{D}|=80K$ &  -14.32 & -13.40 & -11.84 & -8.10 & -14.55\\\cline{1-6}
 \# Parameters & 333K & 267K & 5.7K & 541K & 134M\\
\hline
\end{tabular}\color{black}}
\end{table}

Table \ref{extendedperf} shows that the proposed NN-based approach with an attention module outperforms extended LMMSE when less data is available for training. Even when trained with 80K samples, the proposed NN+Attention approach achieves a reconstruction NMSE very close to the extended LMMSE while reducing the parameter complexity by a factor of $\frac{134M}{333K}\sim 400$, learning to exploit the most significant long range correlations in the channel matrix.
\color{black}


\color{black}
\subsection{Computational complexity}
Our proposed NN-based channel estimation method is trained offline. The training complexity of various methods is compared in terms of the number of trainable parameters for them. In Table \ref{Complexity}, we compared the number of real-valued trainable parameters for various methods in terms of the $L, N$ and $M$ values. Table VI includes a numerical comparison for $L=16, N=32, M=256$ and shows that the proposed NN+Attention approach achieves a reconstruction NMSE very close to the extended LMMSE while reducing its parameter complexity by a factor of $\frac{134M}{333K}\sim 400$. Offline training of our proposed NN+attention channel estimation approach for $L=16, N=32$ and $M=256$ up to 110000 steps takes 2 hours on a NVIDIA GEFORCE RTX 2080 Ti GPU.

After the network is trained, it is deployed for channel estimation. The computational complexity of various methods during deployment is compared in terms of the number of floating point operations (FLOPs) required for channel estimation in each method. The computational complexity for the forward path of our proposed NN-based scheme includes the number of FLOPs executed in the fully connected expansion layers, the convolutional layers and the attention module. The number of FLOPS executed for the fully connected expansion layers is $\mathcal{O}(LNM)$. The number of FLOPs required for each convolutional layer is $\mathcal{O}(NM\rho^2st)$, where $s$ and $t$ are the number of input and output channels for the layer (e.g., $s=t=128$ here) and $\rho^2$ is the kernel size (e.g. $\rho^2=5\times 5, 3\times 3$, etc.). Finally the number of FLOPs required for the matrix multiplications in the attention module is $\mathcal{O}(N^2M^2q)$, where $q$ is the number of output channels in convolutions (we used $q=128$ in simulations). Table \ref{FLOPs} compares the number of FLOPs required for various channel estimation techniques in terms of the channel dimensions $N, M$ and $L$. According to this table, the ``NN+attention" and extended LMMSE methods require the same order of FLOPs while without the attention module, the NN-based method significantly reduces the number of FLOPs required. \color{black}

\begin{table}[!t] 
\centering 
\caption{\color{black}FLOPs for various channel estimators.}\label{FLOPs}
\resizebox{10cm}{!}{%
\color{black}\begin{tabular}{c|c}
   & FLOPs  \\
\hline 
 DP+Attention &  $\mathcal{O}(LNM)+\mathcal{O}(NM\rho^2st)+\mathcal{O}(N^2M^2q)$\\\cline{1-2}
 
 DP &  $\mathcal{O}(LNM)+\mathcal{O}(NM\rho^2st)$\\\cline{1-2}
 
 SP &  $\mathcal{O}(LNM)+\mathcal{O}(NM\rho^2st)$\\\cline{1-2}
 
 LMMSE &  $\mathcal{O}(LNM)$ \\\cline{1-2}
 
 Extended LMMSE &  $\mathcal{O}(N^2M^2)$\\\cline{1-2}
\hline
\end{tabular}\color{black}}
\end{table}

\subsection{Pilot pruning}

Table \ref{PT} provides a summary of the results obtained by our proposed pilot pruning scheme. In particular, we present the NMSE as a function of the sparsity level for different pilot lengths when SNR=10dB, $N=32$ and $M=256$. The numbers in parentheses show the $\lambda$ values used to achieve the sparsity level in each simulation according to Eq. (\ref{pruning}). We observe that the same $\lambda$ value performs fairly well for a number of settings, e.g., we get good results with $\lambda=10^{-6}$ for all $S \le 50\%$ for $L=8$ and 12. We observed that a well-performing $\lambda$ value can always be found by trying only a few negative exponents of 10, i.e., $\{10^{-3}, 10^{-4}, 10^{-5}, 10^{-6}\}$.

According to Table \ref{PT}, the NMSE value degrades slowly as we increase the sparsity; and hence, the proposed pruning scheme can be used to efficiently reduce the pilot overhead without significantly degrading the channel estimation accuracy. For example, when $L=12$, our pilot pruning scheme can save up to $25\%$ of the time-frequency resources while degrading the NMSE by only $1.07$dB. Note that the number of time-frequency resources occupied by pilots is the same for $L=16, S=25\%$ and $L=12, S=0\%$, while the corresponding NMSE values are $-19.43$dB and $-15.35$dB for these two settings, respectively. Hence, our proposed pruning scheme can efficiently allocate pilots non-uniformly along subcarriers to improve the performance while using the same number of time-frequency resources. 

\begin{table}[!t] 
\centering 
\caption{NMSE (dB) performance achieved for different sparsity levels by the pruning-based pilot reduction technique at SNR=10dB, $N=32, M=256$. Values of $\lambda$ (in Eq. (\ref{pruning})) to achieve the specified sparsity level are given in parentheses.}\label{PT}
\resizebox{16cm}{!}{%
\begin{tabular}{c|c|c|c|c|c|c}
  & $S=0\%$ & $S=5\%$ & $S=10\%$ & $S=15\%$ & $S=20\%$ & $S=25\%$ \\
\hline 
 $L=16$ & $-22.88$  & $-21.13 \, (10^{-6})$ &  $-20.11 \, (10^{-5})$ & $-19.85 \, (10^{-5})$ & $-19.57 \, (10^{-5})$ & $-19.43 \, (10^{-5})$\\
\cline{1-7} 
$L=12 \, (10^{-6})$ & -15.35 & -14.59 &  -14.64 & -14.39 & -14.36 & -14.28\\
\cline{1-7}
 $L=8 \, (10^{-6})$ & -14.66 & -14.23 & -13.78 & -13.66 & -13.55 & -13.34\\
\hline
\end{tabular}}
\end{table}

\color{black}To elaborate further, we compare the following five settings in Table \ref{PA}, all of which use $6 \times 256=1536$ time-frequency pilot resources to estimate the channel over a $70 \times 256$ resource grid:

-Pilot scheme A: A rectangular pilot block of size $6 \times 256$ is used for pilot transmission according to the frame structure depicted in Fig. \ref{RG}.  

-Pilot scheme B: In this scheme, one out of every 4 pilots is omitted periodically from a $8 \times 256$ rectangular block of pilots and saved for data transmission. Note that, we still use proposed NN scheme for pilot optimization and channel estimation, but without pruning. Fig. 9a depicts the pilot mask for scheme B.

-Pilot scheme C: In this scheme, we employ the proposed pruning-based pilot allocation scheme with $L=8$ and $S=25\%$. Fig. 9b depicts the pilot mask optimized by our proposed method for this scheme.   

-Pilot scheme D: In this scheme one out of every 2 pilots is omitted periodically from a $12 \times 256$ rectangular block of pilots. Similarly to scheme B, we keep the pilot allocation mask $M$ fixed during training. Fig. 9c depicts the pilot mask for scheme D.

-Pilot scheme E: In this scheme, we employ our proposed pruning-based pilot allocation scheme with $L=12$, and $S=50\%$. Fig. 9d depicts the pilot mask obtimized by our proposed method for this scheme.

Table \ref{PA} shows considerable improvement by the proposed pruning-based pilot allocation scheme in comparison with periodic and rectangular allocations. For example, by starting from a pilot block of $12 \times 256$ and pruning $50\%$ of the pilots with our proposed scheme, we can improve the NMSE by $2.23 \mathrm{dB}$ in comparison with a $6 \times 256$ pilot block, which has the same pilot overhead. It should be noted that, $L=12$ slightly increases the estimation delay at the receiver as the user cannot estimate the channel until all 12 resource blocks are received over the sub-carriers.
\color{black}

\begin{table}[!t] 
\centering 
\caption{\color{black}Pilot allocation performance by NN pruning at SNR=10dB, $N=32, M=256$.}\label{PA}
\resizebox{16cm}{!}{
\color{black}\begin{tabular}{c|c|c|c|c|c}
  Pilot Scheme & A & B & C & D & E \\
\hline 
 NMSE (dB) & $-12.18 $ & $-12.24$ & $-13.34 \, (10^{-6})$ & $-11.96$ & $-14.41 \, (10^{-6})$\\
\hline
 SER & $1.21\times 10^{-2}$ & $1.19 \times 10^{-2}$ & $1.02 \times 10^{-2}$ & $1.25 \times 10^{-2}$ & $8.9 \times 10^{-3}$\\
\hline
\end{tabular}\color{black}}
\end{table}

\begin{figure}[t!]
\centering
\includegraphics[scale=0.27]{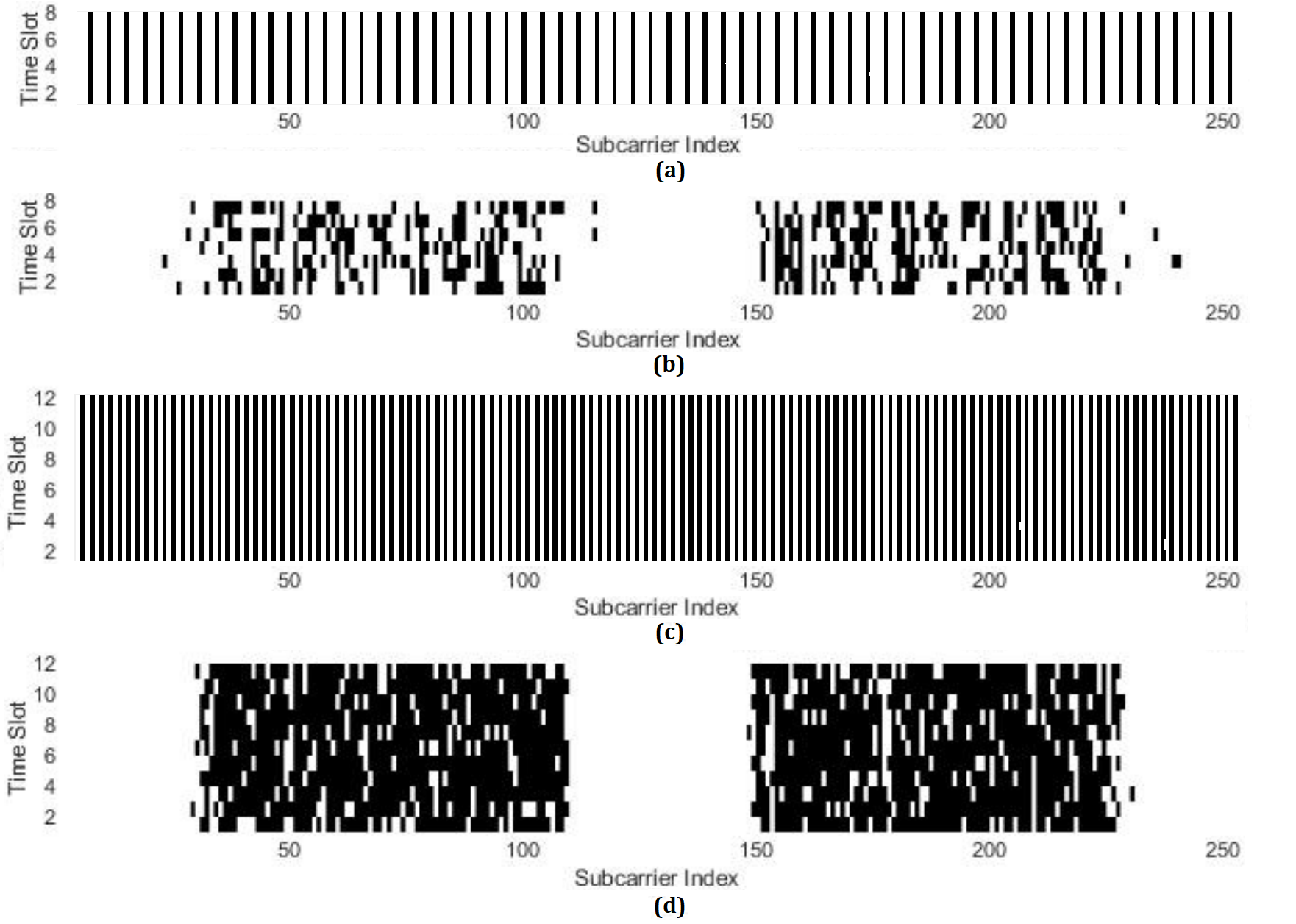}
\caption{\color{black}Illustration of the pilot allocation masks compared in Table \ref{PA}. White dots represent pilots and the black dots represent time-frequency resources saved for data transmission. (a) Periodic removal of pilots with $L=8, S=25\%$, (b) NN optimized pilot pruning with $L=8, S=25\%$, (c) Periodic removal of pilots with $L=12, S=50\%$, (d) NN optimized pilot pruning with $L=12, S=50\%$.}
\label{Alloc}
\end{figure}

Table \ref{PA} shows that when a fixed budget of time-frequency resources is available for pilot transmission, starting from a larger $L$ value and pruning the network down to the available resource budget results in an improved NMSE performance. The NN achieves this by allocating less pilots to those sub-carriers which can be interpolated with sufficient accuracy by the NN exploiting the statistical CSI correlations, while allocating more pilots to other subcarriers. Fig. \ref{Alloc} depicts the allocation masks resulting from the proposed pilot pruning scheme for $L=8, S=25\%$ (Fig. 9b) and $L=12, S=50\%$ (Fig. 9d), where the white dots represent the time-frequency resources devoted to pilots, while the black dots represent those saved for data transmission. As observed in this figure, more resources can be saved for data over transmission subcarriers in the ranges (30-110) and (150-230) in the setting considered.  

\color{black}
In Table \ref{PA}, we also compare the symbol error rate (SER) values for 4-QAM modulation over a $70 \times 256$ time-frequency grid, where the channel is estimated using the five different pilot allocation schemes. We note that we use both the pruned and the data-dedicated time-frequency resources for 4-QAM transmission. For each allocation scheme, we estimate the channel with the trained NN, and use a zero-forcing equalizer over the resource grid, and the maximum likelihood (ML) detector to demodulate the received signal. We report the resulting SER over $10^8$ random symbols. The downlink SNR is 10dB and $N=32, M=256$. Table \ref{PA} shows that the proposed NN-based pilot pruning scheme improves the SER while using the same number of time-frequency resources for pilot transmission; that is, the channel resources saved from the pilot block by the proposed pruning approach can be used for reliable data transmission while reducing the SER.
\color{black}

\section{Conclusion}\label{sec6}
We have proposed a NN-based joint downlink pilot design and channel estimation scheme for FDD massive MIMO-OFDM systems. Our proposed network utilizes dense layers to design pilot signals in a frequency-aware structure followed by convolutional layers which utilize inherent correlations in the channel matrix to provide an accurate channel estimate in an efficient manner. \color{black}We have also employed an attention module to exploit long-range correlations in the channel matrix, which cannot be inferred by the conventional convolutional layers.\color{black} We also proposed an effective pilot reduction technique by gradually pruning less significant neurons from the dense layers to reduce the pilot overhead and to save time-frequency resources for data transmission. Our proposed NN-based pilot design and channel estimation scheme outperforms LMMSE estimation. Moreover, our pruning-based pilot reduction technique effectively reduces the pilot overhead by allocating pilots across subcarriers non-uniformly; allowing less pilot transmissions on subcarriers that can be satisfactorily reconstructed by the subsequent convolutional layers exploiting inter-frequency correlations.

\bibliographystyle{IEEEtran}
\bibliography{refs}
\end{document}